\documentclass[preprint,amsmath,showpacs]{revtex4}
\usepackage{graphicx}
\def\be{\begin{equation}}
\def\ee{\end{equation}}
\def\bee{\begin{eqnarray}}
\def\eee{\end{eqnarray}}

\begin{document}
\title{Intrinsic and oscillated astrophysical
 neutrino flavor ratios revisited}
\author{H. Athar$^{1,}$\footnote{E-mail:
        athar.husain@cern.ch; Present address: 
        Department of Soil, Environmental and Atmospheric 
        Sciences,
         302 ABNR Bldg., University of Missouri-Columbia, 
         Columbia, MO 65211-7250, USA},
        C. S. Kim$^{2,}$\footnote{E-mail:
        cskim@yonsei.ac.kr} and
        Jake Lee$^{2,}$\footnote{E-mail:
         jilee@cskim.yonsei.ac.kr; Present address: 
        Center for Quantum Spacetime, Sogang University,
        Seoul 121-742, Korea.}}
\address{$^{1}$  Physics Division, National
         Center for Theoretical Sciences,
        Hsinchu 300, Taiwan\\
        $^{2}$ Department of Physics, Yonsei
        University, Seoul 120-749, Korea}
\date{\today}
\begin{abstract}
 The $pp$ interactions taking place in the cosmos
around us are a source of
 the astrophysical neutrinos of all the three flavors.
In these interactions,
 the electron and the muon neutrinos mainly come from
the production and the decay
of the $\pi^{\pm}$ mesons, whereas the
tau neutrinos mainly come from the production and the
decay of the $D^{\pm}_{S}$ mesons.
We estimate the three intrinsic neutrino flavor ratios for
 1 GeV $ \leq  E \leq 10^{12}$ GeV in the $pp$ interactions
 and found them to be
 1 : 2 :  $3\times 10^{-5}$.
We study the effects of neutrino oscillations on these
intrinsic ratios. We point out
that the three ratios become 1 : 1 : 1
if $L(\mbox{pc})/E(\mbox{GeV})\geq 10^{-10}$
in the presence of neutrino oscillations,
where $L$ is the distance to the astrophysical
neutrino source in units of parsecs.
\end{abstract}
\pacs{14.60.Lm, 14.60.Pq, 98.70.Sa, 13.15.+g}
\maketitle
\section{Introduction}

Neutrino astronomy holds a  great promise to
explore the interiors of
the dense astrophysical systems, which is not
possible by any other
existing means such as through the study of the
cosmic-rays and/or the
 gamma-rays \cite{12,13,14}.
 It is mainly because once the neutrinos are
produced in the
distant cosmos, they are essentially unobstructed
by the
 intervening background matter
mainly owing to their {\it weak}  interaction
cross section.
 On the other hand, the cosmic-rays (being the
 charged particles)  and the gamma-rays are
either deflected or even considerably absorbed
by the same intervening background matter,
 depending upon the energy \cite{Semikoz:2003wv}.

The presence of the proton component in the
observed cosmic-ray
flux for the entire energy range
(1 GeV $ \leq  E_{p} \leq 10^{12}$ GeV),
 may already be signaling the anticipated
existence  of the neutrino
astronomy.
The search of neutrinos from the
 cosmos may  help to find a unified
explanation for
the common origin of the ultra-high energy
cosmic-rays and the high energy
gamma-rays. The absolute levels of
the astrophysical neutrino
productions are  determined/affected
by the intervening background
 matter \cite{32,33}, the relative
levels nevertheless do not.
 These thus remain important observables
 for the forthcoming detailed astrophysical
neutrino searches.
It is thus important to  investigate the
relative neutrino
production levels in an astrophysical
neutrino source;
as well as  the changes that may occur in the
relative composition of the
neutrino flavors during their propagation to us.

Given the recent empirical evidences of
 the  {\tt neutrino oscillations}
\cite{Maltoni:2004ei},
it is timely to perform a reference study for
the effects of the neutrino oscillations
on the mixed intrinsic
 astrophysical neutrino flavor
ratios during their propagation. Further
motivation is provided by the
 recent developments both in providing
examples to use the
astrophysical neutrinos not only to
study the cosmos
around us \cite{52,53,54,55}, but also to
explore the properties
of the neutrinos itself (including those suggested
beyond the Standard Model of Particle
Physics) \cite{Pakvasa:2004yz}.

We consider the $pp$ interactions as a source
of the intrinsic
 astrophysical neutrino production.
 The first $p$ represent the cosmic-ray flux
produced inside the source,
 whereas the second $p$ represent the medium
contribution in the source.
Commonly cited examples of the astrophysical
systems where the $pp$ interactions
play a role include
the nearby astrophysical sources such as
 the earth atmosphere and the galactic
plane/center region as well as
 other sources inside  our galaxy. The more
 distant  suggested astrophysical neutrino
sources
 include the Active Galactic Nuclei
(AGNs) \cite{Nellen:1992dw},
 and the sites of the Gamma-Ray Bursts
(GRBs) \cite{Dar:2003vf}.

We estimate the three intrinsic neutrino flavor
ratios as a function of the neutrino energy and
study the neutrino oscillation
effects on these for neutrino energy ranging
between 1 GeV and $10^{12}$ GeV.
The {\tt energy dependence} of the three intrinsic
ratios
has not been  studied previously.
We mainly investigate the particle physics
aspects of the
three ratios.
A purpose of the present study is to provide a
firm basis for the relevance of the neutrino
oscillation effects
for the forthcoming searches of the three
astrophysical
neutrino flavor ratios. As in order to determine
 the  astrophysical source characteristics as
precisely and as completely as
  possible via neutrinos, one needs to know the
relevance of the distance to the source $L $
 in the {\tt presence} of the neutrino oscillations
for the given neutrino energy $E$, since the
neutrino
oscillations redistribute the intrinsic neutrino
flavor ratios depending upon the value of the
ratio $L/E$.

This paper is organized as follows.
In Section II, we
estimate the three intrinsic astrophysical
neutrino
flavor ratios in some detail mainly within
the framework of
the Quark-Gluon String Model (QGSM).
In Section III, we study the
effects of the neutrino oscillations on
these. We identify
the range of the $L/E$ values where the
 commonly considered assumption of the
averaging of neutrino oscillation
probabilities may hold.
In Section IV,
we briefly summarize the present status
 of the relevant detection strategies as
well as the detector developments to
search for the
astrophysical neutrino flavor ratios.
In Section V, we present our
conclusions.
\section{The Intrinsic astrophysical neutrino flavor ratios}
Let us define the three intrinsic neutrinos
flavor ratios
 as the $R^{0}_{e/e}$, the $R^{0}_{\mu/e}$,
and the $R^{0}_{\tau/e}$.
Here $R^{0}_{\tau/e}\equiv F^{0}_{\nu_{\tau}}
(E)/F^{0}_{\nu_{e}}(E)$ with
 $F^{0}_{\nu_{\tau}}(E)={\mbox d}
N^{0}_{\nu_{\tau}}/{\mbox d}E$,
for instance. Clearly, $R^{0}_{e/e}=1$.
This ratio
 provides  normalization for  the other
two ratios.
 The previous estimates for
the $R^{0}_{\tau/e}$ ratio are between
 $\geq 10^{-4}$ \cite{Learned:1994wg} and
 $\geq 10^{-5}$ \cite{Athar:1999ym}. However,
the energy
 dependence of the three intrinsic neutrino
flavor ratios was not studied.

We use the following formula for computing
the astrophysical neutrino flux
spectrum
\be
 F^{0}_{\nu}(E)=\frac{\mbox{d}N^{0}_{{\nu}}}
 {\mbox{d}E}=
 d n_{p} \int_{E}^{\infty}
 \mbox{d}E_{p} \; \phi_{p}(E_{p}) \,
 \frac{\mbox{d}\sigma_{pp \to \nu+Y}}
 {\mbox{d}E} \;,
 \label{dNnu}
\ee
where $E$ is the neutrino energy. The
cosmic-ray flux spectrum,
 $\phi_p(E_p)$, is given
by \cite{Mannheim:1998wp}
\be
 \phi_p(E_p)=A (E_p/\mbox{GeV})^{\delta}
 \;\;
 \mbox{cm}^{-2}\mbox{s}^{-1}\mbox{sr}^{-1}
 \mbox{GeV}^{-1},
 \label{cr-flux}
\ee
where $A=0.8$ and $\delta =-2.75$.
We use the above cosmic-ray flux spectrum
for 1 GeV $ \leq  E_{p} \leq 10^{12}$ GeV.
We are aware that the above cosmic-ray flux
spectrum differs by few percent
 for $E_{p}\leq 10 $ GeV
as compared to the more recent
compilation \cite{Gaisser:2002jj}.
 However, this is of not much
concern for our present study since
we are primarily interested
in studying the {\tt astrophysical
neutrino flavor ratios} here.
 Moreover, later, we also study
the effects of varying the  exponent $\delta $.
In the above simplified picture, it is
assumed that all the hadrons and the relevant
leptons decay before interacting with the
medium of the astrophysical neutrino
source.

 The representative proton number density
inside the source is taken to be
 $n_{p}= 1 \;\mbox{cm}^{-3}$, and the
representative
 distance $d$ inside the source is taken
to be $\sim $ 10 kpc,
 where 1 pc $\simeq$ $3 \times
10^{18}$ cm. These values are to merely
represent the
reference absolute levels for the three
neutrino
fluxes. As stated earlier,
 our main concern in this paper is
to study their ratios and their
energy dependence defined
earlier which are obviously independent
of the product $dn_{p}$.
It is clear that the task of computing
the $\mbox{d}N^{0}_{{\nu}}/\mbox{d}E$
 in Eq. (\ref{dNnu}) essentially relies
on the
 evaluation of the differential cross
section $\mbox{d}\sigma/
\mbox{d}E$ in the $pp$ interactions.

In this work, we shall consider only the
most
dominant production channels for each
neutrino flavor generation as the
representative examples.
 Namely, the $\pi^{\pm}$ meson
for the electron and the muon
 neutrinos and
the $D_{s}^{\pm}$ meson for the
tau neutrinos.
 We employ the Quark-Gluon
String Model (QGSM)
to calculate the production
distributions of the above mesons.
 The QGSM approach is non-perturbative
and is based on the
 string fragmentation. It contains a
number of parameters
 determined by the
experiments \cite{132,133,134,135,136}.

 The production cross section of
the hadron $h$ in the QGSM is given by
\be
 \frac{\mbox{d}\sigma^{h} (s,x)}{\mbox{d} x}
 \approx \frac{1}{\sqrt{x^{2}
 + x_\perp^{2}}} \left[\;\sum_{n=1}^{\infty}
 \; \sigma^{pp}_{n} (s) \, \phi^{h}_{n} (s,x)
               +\sigma_{DD}(s)\phi^{h}_{0}(s,x)
 \;\right],
 \label{qgsm}
\ee
where $x=2 p^{h}_{\parallel}/\sqrt{s}$
and $x_\perp = 2\sqrt{ (m^2_{h} +
p^{h2}_\perp)/s}$. The $p^h_\parallel$
($p^{h}_\perp$) is the
 parallel (perpendicular)
momentum of the secondary hadron $h$
in the center
 of mass frame. All the related formulas
are provided in the Appendix.

We have used the VEGAS multi-dimensional
Monte Carlo integration
program \cite{142,143}
 to obtain  the energy distribution
of the neutrino
flux by selecting the events falling
inside the considered energy segment
through a
series of boosts to the laboratory frame.
We  have simplified the secondary
decays of the charged leptons, the $\mu $
and the $\tau $,
 as effectively the two body ones, that
is, a neutrino plus a particle with varying
invariant mass. For instance, we assume
that the $\tau $ lepton decays into a
$\nu_{\tau} $ and a particle $Y$ with the mass
$m_{Y}$ satisfying 0.1 GeV $ < m_Y < m_\tau - 0.1$ GeV.
In our setting $s\sim 2m_{p}E_{p}$. We take the
 $D^{+}_{S}\to \tau^{+}\nu_{\tau}$
branching ratio as $\sim 0.064$
\cite{Eidelman:2004wy}.

We have calculated the following processes
in the QGSM:
\bee
  pp &\to & \pi^\pm (\to \mu\;\nu_\mu) + X
 \nonumber\\
     && \;\;\;\;\;\;\;\;\;\;\to e\;\nu_e
 \;\nu_\mu, \nonumber\\
  pp &\to & D_{s}^{\pm} (\to \tau\;\nu_\tau)
 + X \nonumber\\
     && \;\;\;\;\;\;\;\;\;\;\;\;  \to
 \nu_\tau\; +Y.
 \label{no-label}
\eee

The tau neutrino production via the $D_{s}^{\pm}$
meson can also be calculated in the perturbative
 Quantum Chromo Dynamics (pQCD)
 through $pp\to c\bar{c} \to D^{\pm}_s + X \to
\nu_\tau + Y$. However, the electron and the muon
  neutrino production through the
$\pi^\pm$ meson can not be handled in the pQCD.
 The reason is as follows. There are
many quark-level sub-processes (the $t$-, the $u$-
 as well as the $s$-channel) for the $\pi^{\pm}$
production in the $pp$ interactions, differently
from the $D^{\pm}_{s}$ production which has only the
$c\bar{c}$ pair production channel. Without
some cuts such as on the  $p_T$ or some cuts on the
factorization scale etc., the light quark
productions may blow up as they approach the
non-perturbative region.

In the pQCD calculations, we use the leading-order
 results of the parton sub processes
$q\bar{q},gg \to Q\bar{Q}$ where $Q = c$ for
$pp\to (c\bar{c} \to D^{\pm}_{s} \to )\, \,
\nu_\tau Y$ [and $Q = t$ for $pp\to
t\bar{t}\to \nu X$].
We use a $K$ factor, $K=2$, to account
for the NLO corrections.
For the parton distribution functions,
we use the  CTEQv6 \cite{162,163}.
 We use $m_c=1.35 \mbox{ GeV}$, $m_t=175
\mbox{ GeV}$, $\alpha_s$ with
$\alpha_s(M_Z^2)=0.118$
and $Q^2=\hat{s}/4$ as a factorization scale.
For the $\nu_\tau$ production through
the $c\bar{c}$,
we use the Peterson fragmentation function
with $\epsilon \approx 0.029$
for fragmentation of $c$ or $\bar{c}$ into
the $D^{\pm }_s$ meson \cite{172,173}.

\begin{figure}
 \includegraphics[width=6.75in]{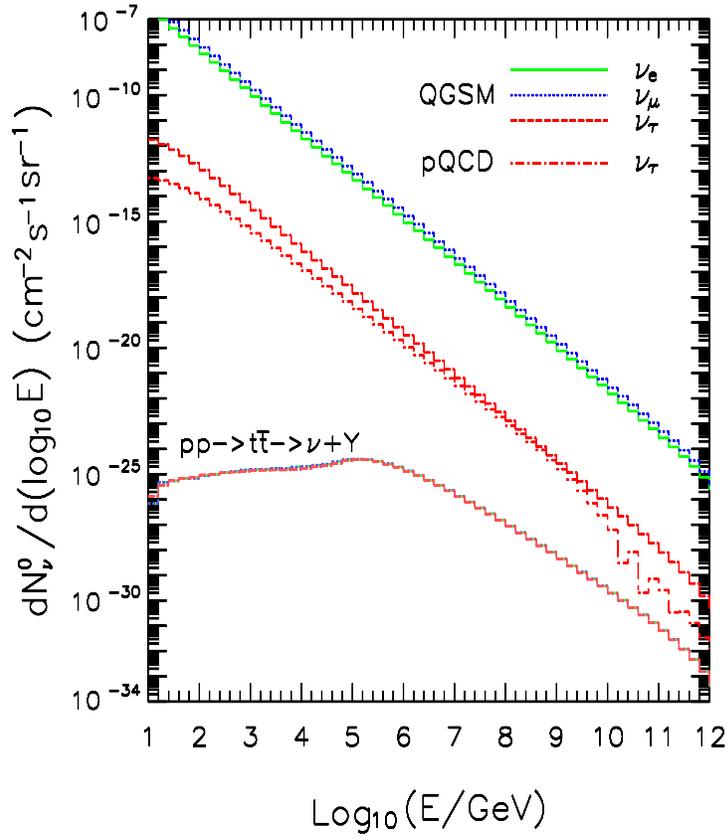}
 \vspace*{-6.0cm}
 \caption{\label{Fig1}The three neutrino
 fluxes
  in the QGSM in the $pp$ interactions
 as a function of the neutrino energy $E$.
    For the tau neutrino, the pQCD result is
 also presented for comparison. The three
 neutrino fluxes in the $pp\to
 t\bar{t}\to \nu + X$, where $t$ is the top quark,
 calculated in the pQCD, are also
 shown as a function of the neutrino energy $E$.}
\end{figure}

Fig. \ref{Fig1} shows the three intrinsic
neutrino flavor fluxes
 as a function of the neutrino energy
 $E$. We plot $\mbox{d}N^{0}_{{\nu}}/
\mbox{d(log}_{10}E)$ in units of
 cm$^{-2}$s$^{-1}$sr$^{-1}$ as a
function of the $E$.
Since the tau neutrino production
via the $D^{\pm}_{S}$
 can be dealt with in both the pQCD
and the QGSM,
 the pQCD
result for the tau neutrinos is also
presented for comparison.
 We note that there is a
relatively large discrepancy at
higher energy ($E\geq 10^{9}$ GeV)
 between the two approaches for
the $\nu_{\tau}$ production.

To consider an example of the
process that may produce the three
 neutrino flavors
without hadronizing in the $pp$
interactions,
 we have studied the $pp \to t\bar{t}$
channel. Here,
the neutrino energy distributions are
not effected
 by the hadronization process. The neutrino
production in the
  $pp\to
t\bar{t}\to \nu X$  is reliably
calculable in the pQCD.
 Here, we consider the direct decays
of $t$
into each lepton (with $\sim 10\%$
branching ratio) and
 include the secondary decays
of massive leptons ($\mu$ and $\tau$).
 Note that all the lepton masses are
very small as
compared to the top quark mass,
as a result all the three neutrino
distributions are of the same
orders of magnitude.
 These results are also
shown in the Fig. \ref{Fig1}.
It is clear from the figure
that the three neutrino {\tt fluxes}
are comparable
 for this process. Compared to
the $\pi^{\pm}$ and the
$D^{\pm}_{S}$ results,
the $t\bar{t}$ contributions are
negligible, less than a factor of
at least $\sim 10^{-3}$
 over the whole range of the
considered energies.

 It can also be seen from the
 figure that
 the production of the $\pi^{\pm}$,
the $D^{\pm}_{s}$,  and their branching
ratios of relevant leptonic decays
are more important than the effects of the lepton
masses.
Our results are in good agreement
for the  $\nu_{e}$ and the  $\nu_{\mu}$
 production with those given in the
Ref. \cite{Ingelman:1996md}
 using the PYTHIA, whereas for the
$\nu_{\tau}$ production,
our results are in good agreement
 with those given in  the
Ref. \cite{Athar:2001jw}.

\begin{figure}
 \includegraphics[width=6.75in]{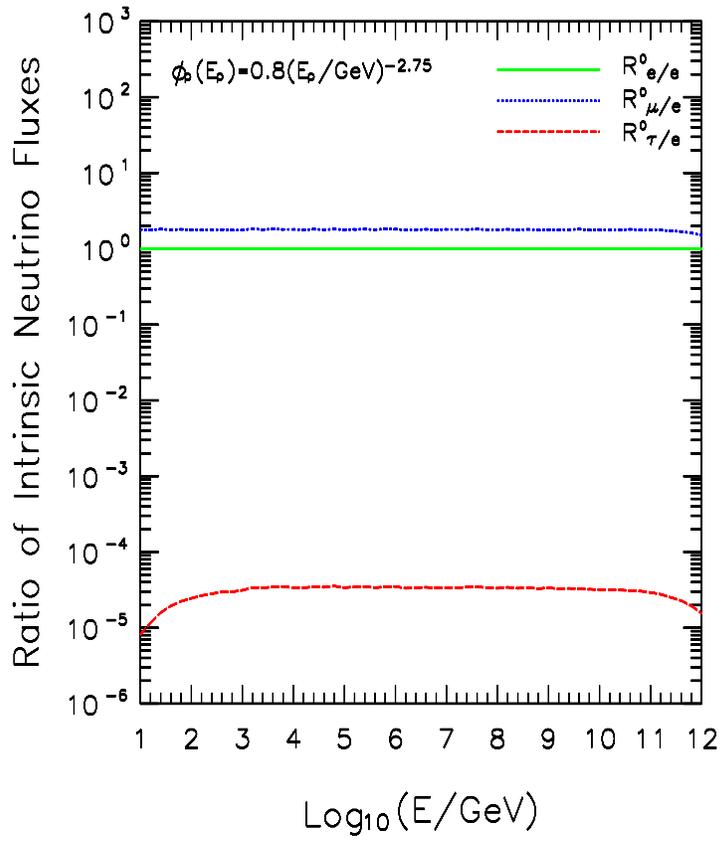}
 \vspace*{-6.0cm}
 \caption{\label{Fig2}The intrinsic
 astrophysical neutrino flavor
 ratios as a function of the neutrino
 energy $E$ in
 the $pp$ interactions. More details
 are given in the text.}
\end{figure}

Fig. \ref{Fig2} shows the three intrinsic
neutrino flavor ratios defined earlier,  as
a function of the neutrino energy $E$.
 We note that an {\tt energy independent}
 relative flux hierarchy among the three
intrinsic neutrino flavor ratios persisted
even at the highest considered energy,
namely
 $R^{0}_{e/e}:R^{0}_{\mu /e}:R^{0}_{\tau /e} \propto
1:2:3\times 10^{-5}$
 for 1 GeV $ \leq  E \leq 10^{12}$ GeV.
 Our results are however subject to the
uncertainties
of extrapolating the parameters of the
hadron/quark production models especially
for $E\geq 10^{6}$ GeV. This corresponds
to center-of-mass energy of
$\sqrt{s}\geq 10^{3}$ GeV. To our knowledge,
the QGSM parameters are fitted up to
 $\sqrt{s} \sim $ 540 GeV using the SPS collider
data for the light mesons, whereas for the
charmed mesons the comparison is
available using the $\sqrt{s} \sim $ 630 GeV data.

\begin{figure}
 \includegraphics[width=6.75in]{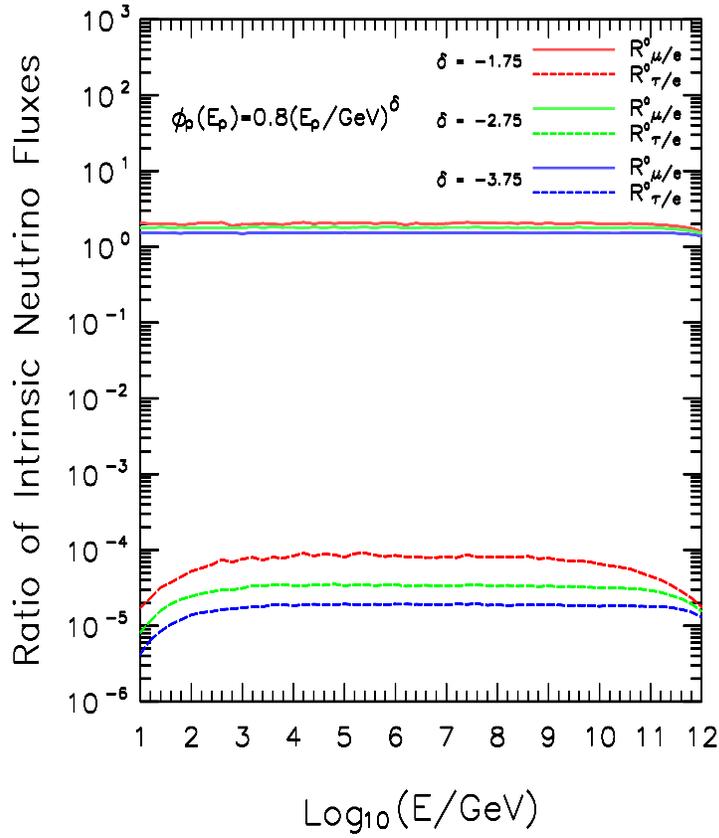}
 \vspace*{-6.0cm}
 \caption{\label{Fig3}The ratios of the
 intrinsic
  neutrino fluxes, the $R^{0}_{\mu/e}$ and
 the $R^{0}_{\tau/e}$,
   for three different cosmic-ray flux
 spectrum
  exponents as a function of the
 neutrino energy $E$.
  The two intrinsic astrophysical
 neutrino flavor ratios
  are essentially independent of the
 cosmic-ray flux spectrum exponent.
   The $\delta =-2.75$
  case is also shown for comparison.
  More details are provided in the text.}
\end{figure}

In Fig. \ref{Fig3}, we show the energy
dependence of the two intrinsic
 ratios of the astrophysical neutrino
fluxes, the
$R^{0}_{\mu/e}$ and the $R^{0}_{\tau/e}$, by
varying the exponent $\delta $
of the cosmic-ray flux spectrum,
$\phi_p(E_p)$ [see Eq. (\ref{cr-flux})],
 in the range $-1.75 \leq \delta \leq -3.75$ .
It is so because
the  cosmic-ray flux spectrum in an astrophysical
source is expected to be harder than the observed
 one locally \cite{Mannheim:1998wp}.
 We note that the ratio $R^{0}_{\tau/e}$ changes
  from $\sim 3\times 10^{-5}$ to
$\sim 8\times 10^{-5}$,
  when the $\delta $ changes from $-2.75$ to $-1.75$
 for 1 GeV $ \leq  E \leq 10^{12}$ GeV.
  However, except the above (slight) change,
 in general, we note from the figure that
the two intrinsic neutrino flavor
ratios are essentially
stable w.r.t. the $\delta $  variation.

For an example of possible energy dependence in the
the three astrophysical neutrino flavor 
ratios coming from astrophysical reasonings, 
 see  \cite{Kashti:2005qa}.
\section{The oscillated astrophysical neutrino flavor ratios}
In this Section, we shall perform a
three neutrino oscillation
analysis of the three intrinsic neutrino
flavor ratios estimated in the
previous Section.
In the context of the three neutrino
flavors, there are 6 independent
neutrino mixing parameters.
The matter effects are found to be
negligible for the
entire range of the neutrino mixing
parameters and the
$E$ values under
discussion \cite{212,213}.

In this analysis, we do not
make the assumption of averaging
over the neutrino oscillation
probabilities.
The neutrino oscillation
effects for the astrophysical neutrinos,
 using the averaged oscillation
probability expressions,
 were studied in some
 detail in Ref. \cite{Athar:2000yw}.
 Here, instead, we shall determine
the $L/E$ range that may be relevant for
the averaging.

We start with the connection $U$
between the flavor $\mid \nu_{\alpha}\rangle $
 and the mass $\mid \nu_{i}\rangle$ eigenstates of the neutrinos, namely
\be
 \mid \nu_{\alpha}\rangle =\sum^{3}_{i = 1}
 U_{\alpha i}\mid \nu_{i}\rangle ,
 \label{m-f}
\ee
 where $\alpha =e, \mu, \tau$. In the
context of the three neutrinos, $U$ is
called the Maki Nakagawa Sakita (MNS)
mixing matrix \cite{Maki:1962mu}.
 Under the assumption that the CP
violating phase  $\delta_{CP} =0 $, the 3$\times$3
 MNS mixing  matrix $U$ in the standard
parameterization connecting
 the neutrino mass and the flavor
eigenstates reads \cite{Eidelman:2004wy}:
\be
 U=\left( \begin{array}{ccc}
          c_{12}c_{13} & s_{12}c_{13} & s_{13}\\
          -s_{12}c_{23}-c_{12}s_{23}s_{13} &
          c_{12}c_{23}-s_{12}s_{23}s_{13} &
          s_{23}c_{13}\\
          s_{12}s_{23}-c_{12}c_{23}s_{13} &
          -c_{12}s_{23}-s_{12}c_{23}s_{13} &
          c_{23}c_{13}
          \end{array}
   \right),
\label{MNS}
\ee
where $c_{ij}\equiv \cos\theta_{ij}$ and
$s_{ij}\equiv \sin\theta_{ij}$. The
presently available
empirical constraints for the
various neutrino mixing parameters,
in the context of the three
neutrino mixing, give
 the following best fit values:
$\theta_{12}=33.2^{\circ}, \, \,
\theta_{23}=45.0^{\circ}, \, \,
\theta_{13}=0.0^{\circ},
 \, \, \delta m^{2}_{12} = 7.9
\times 10^{-5} \, \, {\rm eV}^{2}$ and
    $\delta m^{2}_{23} = 2.1 \times
10^{-3} \, \, {\mbox eV}^2 $ \cite{Maltoni:2004ei}.

Using Eq. (\ref{MNS}),
 the neutrino oscillation probability formula
is \cite{Athar:2003gw}
\be
 P(\nu_{\alpha} \to \nu_{\beta} ;L, E)
 \equiv P_{\alpha \beta}(L, E)=\sum^{3}_{i=1}
 U_{\alpha i}^{2}U_{\beta i}^{2}+
 \sum_{i\neq j}U_{\alpha i}U_{\beta i}
 U_{\alpha j}U_{\beta j}
 \cos \left(\frac{2L}{L_{ij}}\right),
 \label{prob}
\ee
where $\beta =e, \mu, \tau$ and
$L_{ij}\simeq 4E/\delta m^{2}_{ij}$
 is the neutrino oscillation length.
 The $L$  in Eq. (\ref{prob}) is the
 neutrino flight length.

 The neutrino flux ratios
 $R_{\alpha /\beta}(L,E)$,
arriving at the detector,
 in the presence of neutrino
oscillations are estimated using
the relation
\be
 F_{\nu_{\alpha }}(L,E)=\sum_{\beta}
 P_{\alpha \beta}(L,E)
 F^{0}_{\nu_{\beta }},
 \label{tot}
\ee
and the definition
$R_{\alpha /\beta }=
 F_{\nu_{\alpha}}(E)/F_{\nu_{\beta}}(E)$,
 so that
\be
 R_{\alpha /\beta }(L,E)=
 \frac{P_{\alpha e}(L,E)
 +P_{\alpha \mu}(L,E)R^{0}_{\mu /e}
 +P_{\alpha \tau}(L,E)R^{0}_{\tau /e}}
 {P_{\beta e}(L,E)
 +P_{\beta \mu}(L,E)R^{0}_{\mu /e}
 +P_{\beta \tau}(L,E)R^{0}_{\tau /e}},
 \label{R}
\ee
where the $R^{0}_{\alpha /\beta }$
 are taken according to the
 discussion in the previous Section.
The $P_{\alpha \beta}(L,E)$ is
 obtainable
 using Eq. (\ref{prob}). The
unitarity conditions such as
$1-P_{ee}(L,E)=P_{e\mu}(L,E)+P_{e\tau}(L,E)$ are
implemented at each $L$ and $E $
at which these are evaluated.

\begin{figure}
 \includegraphics[width=6.75in]{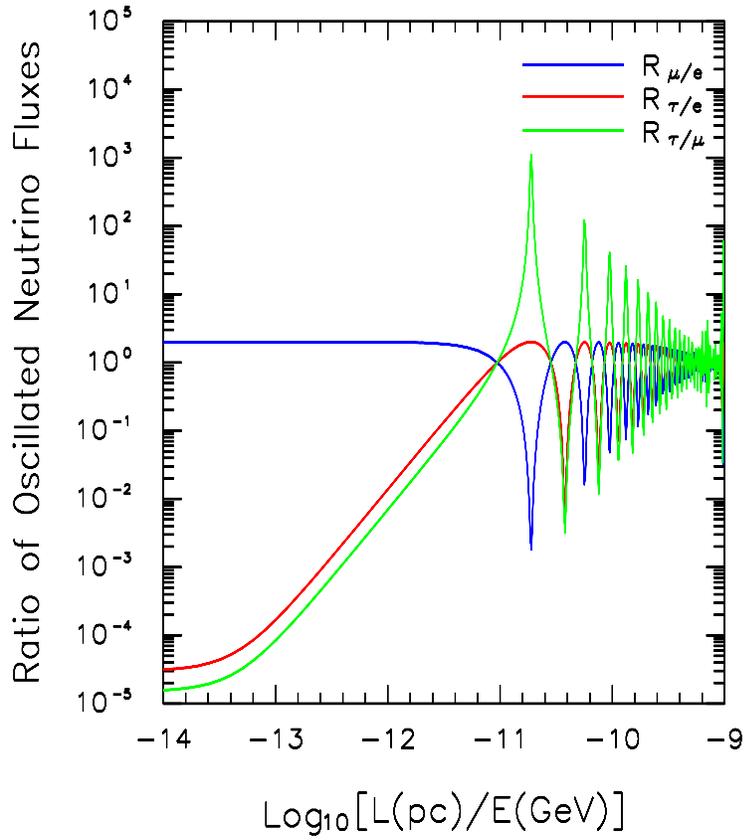}
 \vspace*{-6.0cm}
 \caption{\label{Fig4}The oscillated
 astrophysical neutrino flavor
 ratios as a function of the neutrino
 energy $E$.
 For $L({\rm pc})/E({\rm GeV})
 \geq 10^{-10}$,
 the three ratios enter into a region
 of the relatively rapid and small
 amplitude oscillations centered
 at the abscissa axis value 1.}
\end{figure}

Fig. \ref{Fig4} shows the three
oscillated neutrino flavor
ratios as a function of the
ratio $L/E$. Here,
we have used the fact that the
three intrinsic ratios are
 essentially independent
of the $L$ as well as the $E$.  Note
that the ratio $R_{\tau /\mu}$
behaves  like the ratio $R_{\tau /e}$ .
The ratio
$R_{e/e}$ is not plotted as it is
unaffected by the neutrinos
 oscillations. From the figure, it
is clear that
 the averaging may be a good
approximation
if $L({\rm pc})/E({\rm GeV})
\geq 10^{-10}$, namely
\be
 1:2 :3\times 10^{-5}\longrightarrow
 \nu \, \, \, {\rm osc \, \, \,
 and \, \, \, if}\, \, \,
 L(\rm pc)/E(\rm GeV)\geq
 10^{-10}\longrightarrow
 1: 1:1.
 \label{ratios}
\ee

\begin{figure}
 \includegraphics[width=6.75in]{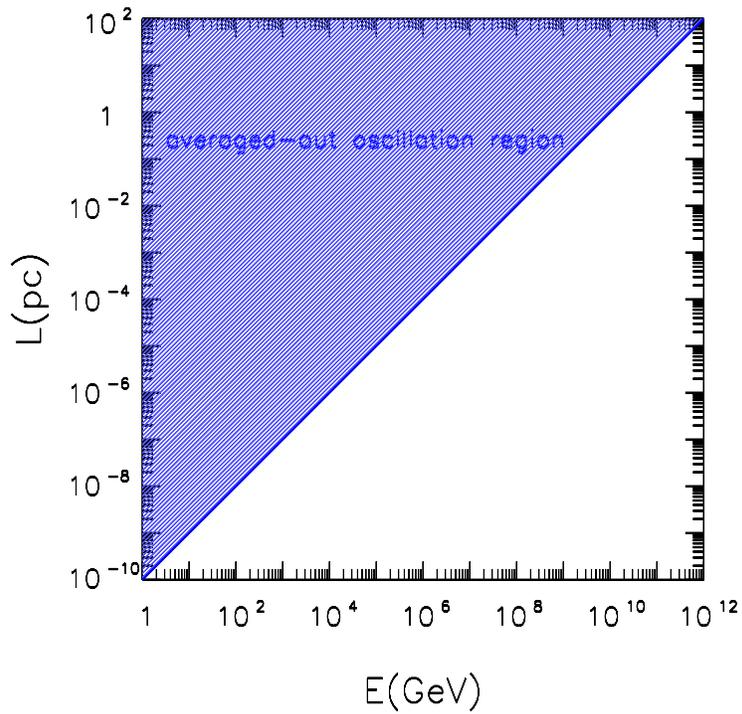}
 \vspace*{-6.0cm}
 \caption{\label{Fig5}The distance
 to the astrophysical
 neutrino source $L$ in units of
 parsecs as a function of the
 neutrino energy $E$ in GeV, using
 the criterion
 $L({\rm pc})/E({\rm GeV})\geq 10^{-10}$.
 The shaded region indicates the
 $L$ and the $E$ value range
 for which the astrophysical
 neutrino flavor
 ratios may be 1:1:1 originating 
 in the 
 $pp$ interactions.}
\end{figure}

Fig. \ref{Fig5}, which is our 
 main result,  shows the region
in the $L$ versus $E$ plane where the
averaging of the neutrino
oscillations probabilities
may be assumed,
 under the above criterion.
For instance, if $E\sim 10^{4}$ GeV,
and if the distance to the source
is $\geq 10^{12}$ cm then the
incoming astrophysical neutrino flux
should be essentially an equal
admixture of the three neutrino flavors.
\section{Prospects for astrophysical neutrino flavor identification}
In this Section, we shall briefly
summarize the presently envisaged
detection strategies and the detector
configurations for the possible
 astrophysical neutrino
flavor identification for
1 GeV $ \leq  E \leq 10^{12}$ GeV
mainly in the context of the
 Cherenkov radiation detection.
For a discussion of
other alternative detection strategies
  such as using the radio and the
acoustic signals,
 see Ref. \cite{252,253}.

 From the reference estimates
presented in Section II,
it follows that
the flux of the intrinsic
astrophysical tau neutrino
flavor is considerably
suppressed for 1 GeV $
\leq  E \leq 10^{12}$ GeV relative
 to that of the electron
and muon neutrino flavor.
  According to the discussion
in Section III,
 the  neutrino oscillations
populate the astrophysical
tau neutrino flux
{\tt comparable} to the
 astrophysical electron
and the
 muon neutrino fluxes,
provided the ratio $L/E$
 is in a  certain range.
 A signature of the neutrino
oscillations in the mixed
astrophysical
neutrino flux  thus shall
be the
 {\tt identification of
the astrophysical tau
neutrino flavor}
among the other two. A
commonly used essential
ingredient
in this context is to make
a comparative  use of the
characteristic
energy dependent astrophysical
tau neutrino induced tau lepton
decay and/or interaction length
scale relative to the relevant
 electron and the muon length
scales \cite{262,263,264}.

\subsection{Current detection strategies}
 The astrophysical neutrino
detection can be achieved
in the neutrino nucleon/electron
 interactions \cite{272,273}.
These interactions
 may occur near or inside the
detector. The charged
 leptons, the (air) showers,
and the associated radiations
 such as the radio and/or
acoustic signals
 are the measurable quantities.
The detectors are optimized
for their performance in
 discriminating the three
neutrino flavors in
 certain energy intervals
\cite{Beacom:2003nh}.
 In addition to reconstructing
the astrophysical
neutrino flavor ratios from an
individual detector,
comparing the data from the
various detectors
 shall also lead
  to the possibility of
identification of  the
three neutrino flavor ratios.
For a recent discussion of
the event
 rates in some representative
astrophysical neutrino
flux models, see
 Ref. \cite{Giesel:2003hj}.
   The astrophysical
neutrinos arrive at an
earth based
 detector in the three
general directions.

 The downward going
astrophysical neutrinos
do not traverse
 any significant cord
length inside the earth
to reach the
detector. In fact, for
large earth surface shower
detectors, the
neutrino nucleon interactions
take place in the
earth atmosphere and the
resulting
 shower (or part of it)
is observed by
the detectors. Several
 studies were performed
to identify
the detector specifications
and/or the energy ranges in which
a specific detector
configuration is/can be optimized to
disentangle the astrophysical
tau neutrino flavor from the
 astrophysical electron and/or
muon neutrino
 flavor through the {\it double
shower} or {\it single shower}
 event topologies
\cite{Learned:1994wg,302,303,304,305}.

The upward going neutrinos
traverse a large earth cord
before they reach the detector.
 At energies $E_{0}\geq
5\times 10^{4}$ GeV,
 the charged current
neutrino nucleon interaction
length is smaller than
the earth diameter.
 As a result significant
{\tt neutrino flavor dependent
 absorption} takes place
 for $E \geq E_{0}$.
 For a discussion of the
upward going
 tau neutrino flavor behavior
versus the upward going
muon neutrino flavor, see Ref.
 \cite{312,313}.

 It might also be possible
to search for the air
 showers induced by
 the incoming (quasi-horizontal)
 astrophysical neutrinos,
in case the
 neutrinos happen to have
just one interaction inside the
 earth. This strategy is
referred to as the earth skimming
 \cite{Feng:2001ue}.
 In some configurations,
the astrophysical neutrinos may
interact just below (within
$5^{\circ}\sim 10^{\circ}$ of) the
  detector horizon
\cite{332,333,334,335,336,337,338}.
The air showers produced
by the earth skimming
 astrophysical neutrinos
can also be searched by the
forthcoming large scale
 balloon/space based detectors
as well \cite{Fargion:2003kn}.
\subsection{Present and the forthcoming detectors}
Presently operating detectors
searching for the
astrophysical neutrinos include the
 Antarctic Muon and Neutrino
Detector Array (AMANDA), its
proposed extension, the Ice
cube \cite{352,353,354}, and  the lake
 Baikal detector
\cite{Aynutdinov:2005hx}. These
detectors use the ice
and the water as detection
 medium respectively, and
are sensitive to all the three neutrino
flavors essentially for
$10^{3}$ GeV $ \leq  E \leq 10^{6}$ GeV
 and mainly search for the upward going (and
horizontal) neutrinos.
 Other under construction
large scale detectors
include the Astronomy with
a Neutrino Telescope
 and Abyss environmental
REsearch (ANTARES)
 project \cite{Sokalski:2005sf}.
For
 $E\leq 10^{3}$ GeV, the
Superkamiokande and
the upcoming one Mega
ton class of detectors
 shall be sensitive
to the three neutrino
 flavors \cite{Goodman:2005xg}.

Several attempts are underway
to implement
the  earth skimming strategy
using the mountains as target
for the
 neutrino nucleon/electron
interactions,
 such as the concept study
carried out by the
 Neutrino Telescope (NuTel)
collaboration \cite{Yeh:2004rp}.
This class of detectors
are/shall be essentially
sensitive for  $10^{6}$
GeV $ \leq  E < 10^{8}$ GeV,
 mainly for the tau
neutrino flavor.
 Earth/air/sea skimming
tau neutrino air shower search
for  more wider energy
range, namely for
 $10^{6}$ GeV $ <  E \leq 10^{10}$ GeV, is also
recently
suggested \cite{402,403}.

The under construction large
surface array detectors such as the
Pierre Auger (PA)
observatory \cite{412,413},
and the Telescope Array (TA)
experiment shall also be sensitive to
all the three neutrino flavors
for
$10^{7}$ GeV $ <  E < 10^{11}$
GeV \cite{Sasaki:2002eg}.
Orbiting Wide-field Light
collector space based mission
(OWL) shall be sensitive
to the 
 $\nu_{e}(\bar{\nu_{e}})$ flavor for
$10^{10}$ GeV $ <  E \leq 10^{12}$
GeV
as it shall search for the
atmospheric
fluorescent trail using
earth atmosphere as the
 detection medium  in the
neutrino nucleon
 interactions \cite{Stecker:2004wt}.
  The Extreme Universe Space
Observatory (EUSO)
shall be sensitive to all
the three neutrino flavors
 by detecting the fluorescent
and the Cherenkov
light produced in the air
showers generated by
the neutrino nucleon
interaction occurring in the
earth atmosphere for
$10^{10}$ GeV $< E \leq 10^{12}$
 GeV \cite{Bottai:2005ic}.

The detectors based on the
 alternative techniques
are also taking data. These
 include the Radio Cherenkov
Experiment (RICE) that is sensitive
to the $\nu_{e}(\bar{\nu_{e}})$ flavor 
based on the anticipated radio-wavelength
Cherenkov radiation detection
that shall be
 produced by the neutrino nucleon
interactions
in the polar ice for $10^{7}$
GeV $< E \leq 10^{12}$
 GeV \cite{452,453}.
The high altitude balloon based Antarctic
 Impulsive transient Antenna
(ANITA) experiment
shall search for the ice skimming
$\nu_{e}(\bar{\nu_{e}})$ flavor 
induced coherent radio signals
 for $E\geq 10^{9}$ GeV
\cite{Miocinovic:2005jh}.
 For $E\geq 10^{11}$ GeV, upper
 limits are also provided by the
 the Goldstone Lunar Ultra-high
energy neutrino
 Experiment (GLUE) mainly for
 the $\nu_{e}(\bar{\nu_{e}})$
 flavor,
based on the similar
 detection technique
\cite{Gorham:2003da}.
 For a summary of  upper
limits
based on the alternative
detection methods
 for $ E >10^{9}$ GeV,
see \cite{Nahnhauer:2004tt}.

\section{Conclusions}
We have performed a  reference
estimate of the three
intrinsic astrophysical neutrino
flavor ratios for the neutrino
energy ranging between 1 GeV
and $10^{12}$ GeV
 in the $pp$ interactions
mainly within the
framework of the Quark-Gluon
String Model (QGSM).
We have taken into account
the $\pi^{\pm}$  meson
production for the generation
of the
 $\nu_{e}$ and the $\nu_{\mu }$
neutrino flavors,
  and the $D^{\pm}_{S}$ meson
production for the
 generation of the $\nu_{\tau }$
 neutrino flavor. We have
also studied  the
 $t\bar{t}$ production
using the perturbative
 Quantum Chromo Dynamics (pQCD)
as an example of the process
for the
generation of the three neutrino
flavors {\tt without hadronization}
 in the $pp$ interactions. The
neutrino generation from the
latter channel is found to be
suppressed for the entire
considered energy range.

 We have only
taken into account the proton
component in the observed
 cosmic-ray flux spectrum.
We have also studied the
variation of the cosmic-ray
flux spectrum exponent $\delta $
on the three ratios and found
that the three intrinsic ratios are
essentially independent of the exponent
 for $-1.75 \leq \delta \leq -3.75$.

 The three astrophysical neutrino
flavor ratios are essentially
 {\tt independent of energy}
and are 1 : 2 : 3$\times 10^{-5}$
for 1 GeV $ \leq  E \leq 10^{12}$ GeV.
 Namely,  the relative intrinsic neutrino
flux hierarchy stays the same for the
 entire considered energy range.
Therefore, the intrinsic
 astrophysical tau neutrino flavor
is relatively suppressed
 for the entire considered energy range.
  Our considered
energy range covers the
entire $E$ range of the
observed cosmic-ray flux.

We have studied the
{\tt effects of the neutrino
oscillations} in
 the three neutrino flavor framework
on the three intrinsic ratios,
using the recent best fit values
 of the neutrino mixing parameters.
 The neutrino oscillation effects
depend upon the distance
 to the astrophysical source $L$
for the given neutrino energy $E$.
 For $L({\rm pc})/E({\rm GeV})
\geq 10^{-10}$, the averaging
 of the neutrino oscillation
probabilities may be assumed,
 where the $L$ is in the
units of parsecs.
 Our present estimate is
intended to provide a firm
basis for the relevance of
the {\tt neutrino oscillations
effects}
for the forthcoming search of
 the astrophysical neutrino
flavor ratios by the 
 various detectors.
\section*{Acknowledgements}
The work of H.A. is supported by
the Physics Division of NCTS.
C.S.K. is supported in part by  CHEP-SRC Program and
in part by Grant No. R02-2003-000-10050-0
from BRP of the KOSEF. J.L. is
supported in part by BK21 program of
the Ministry of Education in Korea and
in part by Grant No. F01-2004-000-10292-0 of KOSEF-NSFC International
Collaborative Research Grant.
\appendix
\section{Formulas for the QGSM}
In Eq.~(\ref{qgsm}), the functions
$\sigma_n^{pp}$ and $\phi_n^h(s,x)$ are
defined as follows:
\bee
  \phi_n^{h}(s,x) &=& a^{h} \left [
    F_{qq}^{h}(x_+,n)F_{q}^{h}(x_-,n)+ F_{q}^{h}
  (x_+,n)F_{qq}^{h}(x_-,n)\right.
    \nonumber \\
    &&\;\;\;\left.+ 2(n-1) F_{sea}^{h}
  (x_+,n)F_{sea}^{h}(x_-,n)\right ] \;\;\;\;\;\mbox{for}\;
    n \ge 1,  \nonumber \\
  \phi_0^{h}(s,x) &=& \frac{3}
  {2}a^{h} (\sqrt{x_+}+\sqrt{x_-})
    F_{q}^{h}(x_+,1)F_{q}^{h}(x_-,1)\;, \nonumber
\eee
where $x_\pm = (\sqrt{x^2+x_\perp^2}
\pm x)/2$ and
\bee
 F_{q}^{h}(x,n) &=& \frac{2}{3}
 \int_x^1 \mbox{d}x_1 \; f_p^{u}(x_1,n) \;G_u^{h}
     \left(\frac{x}{x_1}\right)
 + \frac{1}{3} \int_x^1 \mbox{d}x_1
 \;f_p^{d}(x_1,n) \; G_d^{h}
     \left(\frac{x}{x_1} \right), \\
 F_{qq}^{h}(x,n) &=&
 \frac{2}{3} \int_x^1 \mbox{d}x_1
 \;f_p^{ud}(x_1,n) \;G_{ud}^{h}
 \left(
       \frac{x}{x_1} \right )
 + \frac{1}{3} \int_x^1 \mbox{d}x_1
 \;f_p^{uu}(x_1,n) \;G_{uu}^{h} \left(
     \frac{x}{x_1} \right ), \\
 F_{sea}^{h}(x,n) &=& \frac{1}{4 +
 2\delta_s} \Biggr \{
 \int_x^1 \mbox{d}x_1 \;f_p^{u_{sea}}(x_1,n) \;
    \left[ G_{u}^{h} \left(
 \frac{x}{x_1}\right) + G_{\bar{u}}^{h} \left(
          \frac{x}{x_1}\right ) \right] \nonumber\\
 &&+ \int_x^1 \mbox{d}x_1 \;f_p^{d_{sea}}
 (x_1,n) \;
   \left[ G_{d}^{h} \left(\frac{x}{x_1}
 \right) + G_{\bar{d}}^{h} \left(
        \frac{x}{x_1} \right) \right]
 \nonumber \\
 &&+ \delta_s \int_x^1 \mbox{d}x_1
 \;f_p^{s_{sea}}(x_1,n)
 \;\left[ G_{s}^{h} \left(\frac{x}{x_1}
 \right) + G_{\bar{s}}^{h} \left(
       \frac{x}{x_1}\right ) \right]
 \Biggr \}.
\eee
In the above equations, $f_p^{i}(x,n)$'s are the
distribution functions describing the
$n-$Pomeron distribution functions of quarks or
diquarks ($i=u,d,uu, ...$) with a
fraction of energy $x$ from the proton, and
$G_{i}^{h}(z)$'s are the fragmentation
functions of the quark or diquark chain into a
hadron $h$ which carries a fraction
$z$ of its energy.

The list of the $f_p^{i}(x,n)$ is as follows
\bee
  f_p^{u}(x,n) &=&
    \frac{\Gamma(1+n-2\alpha_N)}{\Gamma(1-\alpha_R)
    ~\Gamma(\alpha_R - 2\alpha_N + n)}
    \times x^{-\alpha_R} ~(1-x)^{\alpha_R -
    2\alpha_N + (n-1)}, \nonumber\\
    &=& f_p^{u_{sea}}, \nonumber\\
  f_p^{d}(x,n) &=&
    \frac{\Gamma(2+n-2\alpha_N)}{\Gamma(1-\alpha_R)
   ~\Gamma(\alpha_R - 2\alpha_N + n+1)}
    \times x^{-\alpha_R} ~(1-x)^{\alpha_R -
  2\alpha_N + n}, \nonumber\\
    &=& f_p^{d_{sea}}, \nonumber\\
  f_p^{uu}(x,n) &=&
    \frac{\Gamma(2+n-2\alpha_N)}{\Gamma(-\alpha_R + n)
  ~\Gamma(\alpha_R - 2\alpha_N + 1)}
 \times x^{\alpha_R - 2\alpha_N + 1} ~(1-x)^{-\alpha_R
 + (n-1)}, \nonumber\\
  f_p^{ud}(x,n) &=&
    \frac{\Gamma(1+n-2\alpha_N)}{\Gamma(-\alpha_R + n)
 ~\Gamma(\alpha_R - 2\alpha_N + 2)}
    \times x^{\alpha_R - 2\alpha_N} ~(1-x)^{-\alpha_R +
 (n-1)}, \nonumber\\
  f_p^{s_{sea}}(x,n) &=&
    \frac{\Gamma(1 + n + 2\alpha_R - 2\alpha_N -2
  \alpha_{\phi})}
    {\Gamma(1-\alpha_{\phi}) ~\Gamma(2\alpha_R - 2
  \alpha_N + n -\alpha_{\phi})}
    \times  x^{-\alpha_{\phi}} ~(1-x)^{2 \alpha_R - 2
  \alpha_N + (n-1) -\alpha_{\phi}}, \nonumber
\eee
where $\Gamma $ is the usual Gamma function.

The list of the $G_{i}^{D_s^{\pm}}(z)$  is given by
\bee
  G_{u,\bar{u},d,\bar{d}}^{D_s^{\pm}} (z)
 &=& (1-z)^{\lambda-\alpha_{\psi} + 2 -\alpha_R -
 \alpha_{\phi}},  \nonumber\\
  G_{uu,ud}^{D_s^{\pm}} (z)
 &=& (1-z)^{\lambda-\alpha_{\psi} +
 \alpha_R -2\alpha_N -\alpha_{\phi} +2},
 \nonumber\\
  G_{s}^{D_s^{+}} (z)
    &=& (1-z)^{\lambda-\alpha_{\psi} +
 2(1- \alpha_{\phi})}, \nonumber\\
    &=& G_{\bar{s}}^{D_s^{-}}(z), \nonumber\\
  G_{s}^{D_s^{-}} (z)
    &=& (1-z)^{\lambda-\alpha_{\psi}}
 \times (1 + a_1 ~z^2),  \nonumber\\
    &=& G_{\bar{s}}^{D_s^{+}}(z). \nonumber
\eee

The list of the $G_{i}^{\pi^{\pm}}(z)$  is given by
\bee
  G_{u}^{\pi^+} (z)
 &=& (1-z)^{-\alpha_R + \lambda}  \nonumber\\
 &=& G_{\bar{d}}^{\pi^+} (z) = G_{d}^{\pi^-}
(z) = G_{\bar{u}}^{\pi^-} (z), \nonumber\\
 G_{d}^{\pi^+} (z)
 &=& (1-z)^{-\alpha_R + \lambda + 1}  \nonumber\\
 &=& G_{\bar{u}}^{\pi^+} (z) = G_{u}^{\pi^-} (z) =
 G_{\bar{d}}^{\pi^-} (z), \nonumber\\
 G_{uu}^{\pi^+} (z)
 &=& (1-z)^{\alpha_R - 2\alpha_N + \lambda}
 \nonumber\\
 &=& G_{\bar{ud}}^{\pi^+} (z), \nonumber\\
 G_{uu}^{\pi^-} (z)
 &=& (1-z)^{\alpha_R - 2\alpha_N + \lambda + 1}
 \nonumber\\
 &=& G_{\bar{ud}}^{\pi^-} (z). \nonumber
\eee

In the above, the input parameters are as follows:
\bee
  &&\alpha_R = 0.5 ~,~\alpha_N = -0.5 ~,~\alpha_{\phi} =
 0 ~,~\alpha_{\psi} = -2.18 ~,~\lambda = 0.5 ~,~a_1 =
 5\nonumber\\
  &&\delta_s = 0.25 \; (0), \;\;\; \mbox{if strange
 sea contribution is turned on (off),}
  \nonumber\\
  &&a^{D_s} = 0.0007 ~,~a^{\pi} = 0.44 ~.\nonumber
\eee

The function $\sigma_n^{pp} (s)$ and $\sigma_{DD}(s)$
are given by the following
formulas:
\bee
 \sigma_n^{pp}(\xi) &=& \frac{\sigma_p}{n~z} \;\left
     ( 1-\exp(-z)\sum_{k=0}^{n-1} ~\frac{z^k}{k!}
 \right) \;, \nonumber \\
 \sigma_{DD}(\xi) &=& \frac{C-1}{C}\sigma_p
 [f(z/2)-f(z)] \;, \nonumber
\eee
where
\be
 \xi = \ln \left(\frac{s}{1 ~(\mbox{GeV})^2}
 \right)\;,\; z
 =\frac{2C~\gamma_p}{R^2+{\alpha_p}^{\prime}
 ~\xi} ~\exp(\xi \Delta)\;,\;
 \sigma_p = 8\pi \gamma_p ~\exp(\xi
 \Delta)\;, \nonumber
\ee
and
\be
 f(z)=\sum_{\nu=1}^{\infty}\frac{(-z)^{\nu-1}}
 {\nu\nu !}
 = \frac{1}{z}\int_0^z dx \frac{1-e^{-x}}{x}
 \;.\nonumber
\ee

The best fit parameter values are as follows :

{\noindent (i) for $\sqrt{s} ~\leq~ 10^3$ GeV}
\bee
  &&\gamma_p          = 3.64 ~ ~(\mbox{GeV})
 ^{-2},\;\;
  R^2               = 3.56 ~ ~(\mbox{GeV})^
 {-2},\;\;
  \alpha_p^{\prime} = 0.25 ~ ~(\mbox{GeV})
 ^{-2}, \nonumber \\
  &&C                 = 1.5\;,\;
  \Delta            = 0.07\;. \nonumber
\eee
(ii) for $\sqrt{s} ~\geq~ 10^3$ GeV
\bee
  &&\gamma_p          = 1.77 ~ ~(\mbox{GeV})
 ^{-2},\;\;
  R^2               = 3.18 ~ ~(\mbox{GeV})
 ^{-2},\;\;
  \alpha_p^{\prime} = 0.25 ~ ~(\mbox{GeV})
 ^{-2}, \nonumber \\
  &&C                 = 1.5\;,\;
  \Delta            = 0.139\;. \nonumber
\eee
\pagebreak
\pagebreak
\pagebreak
\pagebreak
\pagebreak
\pagebreak
\end{document}